\documentstyle[aps,pra,preprint,tighten]{revtex}
\begin{document}

\title{\bf $ $ \\
On the Inequivalence of Weak Localization \\
and Coherent Backscattering}

\author{ $ $ \\
Matthew B. Hastings,$^{(1)}$ A. Douglas Stone,$^{(1)}$ and
Harold U. Baranger$^{(2)}$}

\address{ $ $ \\
$^{(1)}$ Applied Physics, Yale University,
P.O. Box 208284, New Haven, CT 06520-8284 }

\address{
$^{(2)}$ AT\&T Bell Laboratories 1D-230, 600 Mountain Avenue,
Murray Hill, NJ 07974-0636 \\
$ $ }

\date{Submitted to Phys. Rev. B, April 1994}

\maketitle

\begin{abstract}
We define a current-conserving approximation for the local conductivity
tensor
of a disordered system which includes the effects of weak localization.
Using this approximation we show that the weak localization effect in
conductance is not obtained simply from the diagram corresponding to
the coherent back-scattering peak observed in optical experiments.
Other diagrams contribute to the effect at the same order and
decrease its value. These diagrams appear to have no
semiclassical analogues, a fact which may have implications
for the semiclassical theory of chaotic systems.  The effects of discrete
symmetries on weak localization in disordered conductors is evaluated and
and compared to results from chaotic scatterers.
\end{abstract}


\narrowtext

\section{Introduction}

The weak localization (WL) effect has played a seminal role in both
theoretical
and experimental work on quantum transport in disordered conductors.
On the theoretical side it enters the scaling theory of localization
in a fundamental manner \cite{gang4,Bergmann}, and on the experimental side it
provides
the best measurement of the phase-coherence length for electrons at low
temperature \cite{Bergmann}.  Finally,
through the normal-metal Aharonov-Bohm effect, it
provides a very direct measurement of quantum interference of normal
electrons on the scale of this phase-coherence length \cite{RMP}.
The quantitative
theory of weak-localization is based on impurity-averaged perturbation
theory in the small parameter $(k_f l)^{-1}$ ($k_f$ is the fermi wavevector
and
$l$ is the elastic mean free path) \cite{AGD}.  Many experiments over the
past
decade have tested this theory in detail and found excellent agreement with
the measured low-temperature variation of the conductance with
temperature and magnetic field.

The weak localization effect arises due to time-reversal (TR) symmetry. It is
often described \cite{Bergmann}
by noting that a diffusing electron wave has an enhanced
probability of returning to its starting point (compared to a classical
particle) due to the constructive interference
of time-reversed pairs of trajectories forming closed loops.  This
enhancement of the return probability then leads
to reduced conductance with respect to the classical Drude value.
Breaking time-reversal symmetry eliminates this interference causing
a positive magnetoconductance (in the simplest case in which spin-orbit
scattering is negligible \cite{Bergmann}).
Although WL is an interference effect and hence
clearly non-classical, it is often described as a {\it semiclassical}
effect \cite{Bergmann,cs,arg}
which can be understood simply by adding the amplitudes for
motion along classical trajectories with appropriate phases, and then
squaring the amplitude.  Since recent theoretical advances in semi-classical
quantum theory have shown quantitative success in other fields, e.g.
atomic physics \cite{Holle};
it is of some interest to see if weak localization can be
quantitatively obtained from a summation over classical paths.  Recently
two of the authors addressed this problem in a treatment of WL in
ballistic conductors and numerically found contributions to the WL effect
which they could not obtain from a semiclassical calculation \cite{wl1,wl2}.
It was noted \cite{wl2}
that a similar contribution apparently entered in the WL theory of
disordered conductors, but had not been explicitly discussed or calculated
in the literature.  We present such a calculation below and discuss its
implications.  Reference  \cite{wl1} also discovered a sensitivity
of WL in ballistic conductors to discrete spatial symmetries; we evaluate
this sensitivity for disordered conductors below.

Well before the discovery of weak localization in disordered conductors
the increase in backscattering probability due to time-reversal symmetry
had been noted both in optics \cite{Watson,deWolf} and in nuclear scattering
 \cite{nucref}.  In the former this effect is referred to
as {\it coherent backscattering} (CB)
and in the latter case as the {\it elastic enhancement factor}.
In optics one clearly observes a peak approaching a factor of two in
height in the backwards direction \cite{van,maret,akker}
(if the polarization is preserved).
Recently the CB effect in optics has been hypothesized \cite{astro}
to explain the
surge in brightness of astronomical bodies when in opposition,
and this hypothesis is believed to have significant implications for
planetary science.  Although known for some time, the detailed study of
CB in optics was stimulated by WL theory and the two
terms (weak localization and coherent backscattering) are often used
interchangeably.  In this work we emphasize that the two effects are not
equivalent: diagrams which make a negligible contribution to the
differential scattering cross-section make a contribution to the
weak localization conductance of the same order and of opposite sign as
the diagram which gives rise to the CB peak.

\section{Current Conservation for Exact Conductivity and Conductance}

In linear response theory the quantum expectation value of the
local DC current density induced by an electric field
${\bf E}({\bf r^{\prime} })$ is
\begin{equation}
 <{\bf J}({\bf r})> =
\int d{\bf r^{\prime} } {\underline \sigma} ({\bf r},{\bf r^{\prime} })\cdot
{\bf E}({\bf r^{\prime} })
\end{equation}
where ${\underline \sigma}$ is the Kubo local conductivity tensor, which may be
expressed
in terms of the exact eigenfunctions and eigenenergies for a given disorder
configuration \cite{economou,fisher,bs}.
The general expression for ${\underline \sigma}$ in the presence of a
magnetic field, $B$, may be divided into a term which only depends on
states at the fermi surface (at T=0) and is symmetric in B, and a term
which depends on all states below the fermi surface and is anti-symmetric
in
B \cite{bs}.  The weak localization effect relates to average properties of
conductivity and conductance which are symmetric in B, hence the
anti-symmetric term may be neglected. With this simplification we may
write
\begin{equation}
 {\underline \sigma} ({\bf r},{\bf r^{\prime} })=
- \frac{e^2 \hbar^3}{16\pi m^2}
\Delta G(\varepsilon_{f},{\bf r},{\bf r^{\prime} })
{\stackrel{\leftrightarrow}{D}}* {\stackrel{\leftrightarrow}{D}}
\Delta G(\varepsilon_{f},{\bf r^{\prime} },{\bf r}),
\end{equation}
where $\Delta G(E,{\bf r},{\bf r^{\prime} })
\equiv {G^+({\bf r},{\bf r^{\prime} })} - {G^-({\bf r},{\bf r^{\prime} })} $,
$G^{\pm}$ is
the advanced (retarded) Green functions for the single particle Schr\"odinger
equation, and ${\stackrel{\leftrightarrow}{D}}$ represents the antisymmetric
gauge-invariant derivative:
$f(x){\stackrel{\leftrightarrow}{D}} g(x)= f(x)[Dg(x)]-g(x)[Df(x)]$,
$D={\bf \nabla} - (ei/\hbar c){\bf A}({\bf r})$.
Since WL occurs at relatively weak magnetic fields it turns out to be
sufficient to make the further approximation $D \approx \nabla$ which we
do henceforth.  With this set of approximations DC current conservation for
the local current density, $ \nabla \cdot <{\bf J}({\bf r})> = 0$, is
equivalent
 \cite{bs} to the condition
$\nabla \cdot {\underline \sigma} = \nabla ' \cdot {\underline \sigma} =0$.

Transport experiments do not directly measure the local conductivity
tensor;
instead they measure conductance or resistance.  In the simplest case of
a two-probe measurement, in which the voltage drop is measured between
the
current source and sink, the conductance, $g$, is just the inverse of the
resistance and is related to the local conductivity by integration over
the cross-sections at the interface between the sample and the
leads \cite{bs}:
\begin{equation}
 g= \int dS_1 \int dS_2 \hat{n}_1 \cdot
{\underline \sigma}({\bf r},{\bf r^{\prime} }) \cdot \hat{n}_2
\end{equation}
where $\hat{n}_1,\hat{n}_2$ are unit normal vectors to the cross-sections
of the two leads.

There are two different forms of Eq. (3) often used in the literature;
the first reexpresses conductance in terms of scattering coefficients, and
the second reexpresses conductance as an average over the sample volume.
In the first case, the cross-sections are taken
just outside the sample in order to express ${\underline \sigma}$ in terms
of the exact asymptotic wavefunctions which involve the transmission and
reflection amplitudes.  Straightforward manipulations worked out by
Economou and Soukolis \cite{economou} and by Fisher and Lee \cite{fisher}
then show that Eq. (3) is equivalent to the two-probe Landauer formula:
\begin{equation}
 g = \frac{e^2}{h} \sum_{i,j}^{N_c} T_{ij}
\end{equation}
where $T_{ij}$ is the transmission coefficient between the propagating
states $i,j$ in the leads (assumed identical) and there are $N_c$ such
states at $\varepsilon_{f}$.  Below we will discuss extensively the
contribution to the WL effect of the transmission coefficients which are
diagonal or off-diagonal in the mode indices $i,j$.

In the second case one uses the relation
$\nabla \cdot {\underline \sigma} = \nabla ' \cdot {\underline \sigma} =0$
to show that g is independent of the location of the cross-section in Eq. (3)
and then integrates over both $x,x'$ to obtain
\begin{equation}
 g= - \frac{e^2 \hbar^3}{16\pi m^2 L_x^2} \int d{\bf r} d{\bf r^{\prime} }
\Delta G(\varepsilon_{f},{\bf r},
{\bf r^{\prime} }){\stackrel{\leftrightarrow}{\nabla}}_x
{\stackrel{\leftrightarrow}{\nabla}}_{x'}\Delta
G(\varepsilon_{f},{\bf r^{\prime} },{\bf r}),
\end{equation}
where the integrations are over the entire sample.  This volume-averaged
form for the conductance has been used in most previous studies of WL;
as we will see it leads to a much simpler set of diagrams for the average
conductance than does Eq. (4).  We note that the volume-averaged form is
only valid for two-probe conductance, and treatments of four-probe
resistance measurements must rely on Eqs. (3)and (4) \cite{bs,Kane1,Kane2}.
This is particularly important for mesoscopic samples, where the lead geometry
may have significant influence on the measurements.

\section{Current-Conserving Approximations for $<{\underline \sigma}>$}

\subsection{Impurity-Averaged Perturbation Theory}

Disordered conductors are typically characterized by their statistical
properties when averaged over an ensemble of impurity configurations.
As noted above, the WL effect appears in the average conductance; the
primary
analytic tool for calculating this effect is impurity-averaged perturbation
theory in the small parameter $(k_fl)^{-1}$.  This perturbation theory
is equivalent to that for electrons with a static interaction, except for
the absence of closed fermion loops (which is explained naturally in the
replica formulation of the problem).  As in any diagrammatic expansion, it
is crucial to evaluate a set of diagrams at each order which maintains the
conservation laws present in the exact theory.  In this case the important
conserved quantity is current, and we seek a set of diagrams for
$<{\underline \sigma} ({\bf r},{\bf r^{\prime} })>$ which both describes the WL
effect and
satisfies
\begin{equation}
\nabla \cdot <{\underline \sigma}> = \nabla' \cdot <{\underline \sigma}> = 0 ,
\end{equation}
where
henceforth both the angle brackets and overbars
will denote the average over impurity configurations.

We shall adopt the standard white-noise model for
the statistics of the impurity potential: $<V({\bf r})> = 0,
<V({\bf r})V({\bf r^{\prime} })>= c_iu^2\delta ({\bf r} - {\bf r^{\prime} })$,
where $c_i$ is the density of
impurities and all higher cumulants are assumed zero.  The usual
approximation (see e.g. Ref. \cite{scot})
for the average Green function
is the self-consistent Born approximation (SCBA)
which is shown diagrammatically in Fig. 1; it may be
expressed for our case by the self-consistent equation:
\begin{equation}
 [E - {c_i u^2} \overline{G}({\bf r},{\bf r}) +
\frac{\nabla^2}{2m}] {\overline{G}({\bf r},{\bf r^{\prime} })} =
\delta({\bf r} - {\bf r^{\prime} }) .
\end{equation}
Although we will consider finite disordered samples connected to ordered
leads (and this will be important for certain aspects of the calculation)
the violation of translational symmetry at the boundaries is negligible for
the average Green function.  This is first because $\overline{G}$
decays exponentially for
$|{\bf r} - {\bf r^{\prime} }| > l $ and is insensitive to the boundaries
over most of the sample, and secondly because even at the boundaries
corrections to the average Green function due to the interface only appear
at next order in $(k_fl)^{-1}$.  Since the boundary effects are found to be
negligible in
Eq. (7) we may treat the system as translationally invariant on average, hence
$\overline{G}({\bf r},{\bf r},E)$ is just a complex function of energy,
independent of ${\bf r}$.  For weak magnetic field even the energy-dependence
is negligible and one finds
${c_i u^2} \overline{G}^{\pm}({\bf r},{\bf r}) = \mp i/2\tau$,
where $\tau$ is the elastic scattering time.  It then follows from Eq. (7)
that
\begin{equation}
\overline{G}^{\pm}({\bf r},{\bf r^{\prime} },E) =
G_0^{\pm}({\bf r},{\bf r^{\prime} },E \pm i/2\tau),
\end{equation}
where $G_0$ is the Green function of the problem in the absence of disorder.
Substituting this relation into the known form of $G_0$ leads to the
exponential decay with $l$ mentioned above.

\subsection{Current Conservation in the Ladder Approximation}

At leading order in $(k_fl)^{-1}$ it is well-known \cite{hersh}
that the SCBA
for the average Green function will yield a current-conserving
approximation
for the conductivity if the velocity vertices are dressed by the infinite
sum of ladder diagrams (see Fig. 2).  This ladder sum
$L({\bf r},{\bf r^{\prime} })$ satisfies the integral equation:
\begin{equation}
 L({\bf r},{\bf r^{\prime} }) = \delta({\bf r}-{\bf r^{\prime} }) +
\int {\bf dr_1} L_0({\bf r},{\bf r_1}) L({\bf r_1},{\bf r^{\prime} })
\end{equation}
where the kernel $L_0({\bf r},{\bf r^{\prime} }) =
{c_i u^2} |{\overline{G}^+} ({\bf r},{\bf r^{\prime} })|^2$.

The proofs of current conservation in various approximations to
$<{\underline \sigma}({\bf r},{\bf r^{\prime} })>$ rely on properties of the
velocity vertex under the divergence operation.  The most general (bare)
velocity vertex which can appear has the form
\begin{equation}
 V_0({\bf r},{\bf r_{1} },{\bf r_{2} }) =
{G^+({\bf r_{1} },{\bf r})}\frac{{\stackrel{\leftrightarrow}{\nabla}}}{2mi}
{G^-({\bf r},{\bf r_{2} })}
\end{equation}
where ${\bf r_{1} },{\bf r_{2} }$ are normally integrated over in evaluating
the diagram and ${\stackrel{\leftrightarrow}{\nabla}}$ is the anti-symmetric
gradient operator.  To prove current conservation for any set of diagrams
we must take the divergence of each diagram
in the chosen set and show that the resulting sum vanishes.  Since the
${\bf r},{\bf r^{\prime} }$ dependence in each diagram resides solely in the
vertices we need only evaluate the divergence of the relevant vertex and
then convolve this with the rest of the diagram.  Following Hershfield
\cite{hersh}, we can regard the
operation of taking the divergence as generating a new set of diagrams
with the velocity vertex replaced by a ``divergence vertex'' which simply
designates the point at which the divergence was taken
(we denote this divergence vertex by a black dot in the figures); the resulting
set of diagrams must sum to zero if the approximation conserves current.

The divergence of the vertex $V_0$ may
be evaluated using Eq. (7) to substitute for the factors
$\nabla^2 {\overline{G}} /2m$ which arises after differentiation.
\begin{equation}
\nabla \cdot V_0({\bf r},{\bf r_{1} },{\bf r_{2} }) =
\frac {1}{\tau} {G^+({\bf r_{1} },{\bf r})} {G^-({\bf r},{\bf r_{2} })} -
i{G^-({\bf r},{\bf r_{2} })} \delta ({\bf r}-{\bf r_{1} })
+i{G^+({\bf r_{1} },{\bf r})} \delta({\bf r}-{\bf r_{2} }).
\end{equation}
The diagrams generated by this procedure are shown in Fig. 3.  We see that
in general taking the divergence of a bare vertex in a conductivity diagram
generates three diagrams: the first and second diagrams have one of the
Green functions deleted and the divergence vertex moved to
${\bf r_{1} },{\bf r_{2} }$ respectively, and the third is identical
to the original conductivity diagram except that the divergence vertex
replaces the velocity vertex and is multiplied by a factor $1/\tau$,

Now consider dressing the bare velocity vertex with an impurity line as
shown in Fig. 4a.  In this case the bare vertex connects to the same
point, ${\bf r_0}$; if we set ${\bf r_{1} } = {\bf r_{2} } ={\bf r_0}$ in
Eq. (11) the second and third terms combine using the identity
$-i{c_i u^2}[{\overline{G}^-}({\bf r},{\bf r}) -
{\overline{G}^+} ({\bf r},{\bf r})]= -1/\tau$ and yield a single
diagram with the impurity line removed as shown in Fig. 4a.  A similar
result occurs when the divergence is taken of a bare vertex
dressed by two impurity lines.
It is evident from the figure that the first diagram generated
from the singly dressed vertex is the negative of the second diagram
generated from the doubly dressed vertex and will cancel if these two
diagrams are summed.
It is very useful at this point to define the dressed vertex $\tilde{V}$
as the vertex $V_0$ dressed with a full ladder sum (note that the bare
vertex appears as the first term of $\tilde{V}$).  Because of the cancellation
of successive diagrams generated by taking the divergence of a velocity
vertex connected to a ladder we obtain the simple result
\begin{equation}
\nabla \cdot \tilde{V}({\bf r},{\bf r_{1} },{\bf r_{2} }) =
 - i{G^-({\bf r},{\bf r_{2} })} \delta ({\bf r}-{\bf r_{1} })
+ i{G^+({\bf r_{1} },{\bf r})} \delta({\bf r}-{\bf r_{2} }).
\end{equation}
This result is shown diagrammatically in Fig. 4b.
The divergence operation on the dressed vertex generates
only two diagrams, one with the point ${\bf r}$ shifted to the point
${\bf r_{1} }$,
the other with it shifted to the point ${\bf r_{2} }$, with the analyticity of
the Green function reversed, and with opposite sign.   These two diagrams
will have the same number of impurity lines.

With this identity it is easy to confirm current conservation in the
ladder approximation for $<{\underline \sigma}>$ (this approximation is
represented diagrammatically in Fig. 5).  Taking the divergence at the dressed
vertex generates two diagrams with a single Green function multiplied
by $\delta ({\bf r} -{\bf r^{\prime} })$ attached to a bare velocity vertex.
The antisymmetric derivative of these diagrams vanishes due to the
symmetry of the Green functions under interchange of
${\bf r},{\bf r^{\prime} }$.  Thus, as claimed, the divergence of
${<\underline \sigma >}$ in this approximation is zero.  It is worth
noting that this argument establishing current conservation remains valid
even at the interface of between the sample and the leads.

\subsection{Current Conservation Including Weak Localization}

The ladder approximation to ${<\underline \sigma >}$ discussed in the last
section does not, however, describe the WL effect: the WL effect only appears
at next order in $(k_fl)^{-1}$. We have not
found in the literature an explicitly current-conserving set of diagrams
for ${<\underline \sigma >}$ which upon integration yields the WL contribution
to the
conductance.  However, it is known \cite{Bergmann} that the volume-averaged
expression for $g$ yields the full effect when one evaluates only the
diagram with one crossed ladder (Fig. 2b) without vertex corrections;
this implies that all other volume-averaged diagrams at this order
either vanish or sum to zero.  Thus the WL effect in the {\it total}
conductance may be calculated without determining
an explicit current-conserving set for $<{\underline \sigma}>$.
However in order to evaluate individual transmission or reflection coefficients
$<T_{ij}>,<R_{ij}>$ and determine their separate contributions to $<g>$,
the identification of
such a set is necessary.  A current-conserving approximation which includes
the WL effect is represented by the diagrams shown in Fig. 6.
This approximation will be shown to conserve current {\it exactly}
(if the SCBA is used for the average Green function)
by a generalization of the
type of argument \cite{hersh} used for the simple ladder approximation to
$<{\underline \sigma}>$ and for conductance fluctuations in multi-lead
structures.  The
set given includes diagrams of higher order in $(k_fl)^{-1}$
than needed for the WL effect; however, inclusion of the higher-order diagrams
actually expedites the proof.  (We will not evaluate all diagrams
explicitly anyway).
In these diagrams the crossed insertion represents the time-reversed or crossed
ladder shown in
Fig. 2b, while the number on each insertion represents the number of
impurity scatterings.  We will show that the divergence of this set
for a fixed total number of impurity lines is zero; summing over these sets
for all numbers of impurity lines in the ladder
yields a complete set of current-conserving diagrams
containing the crossed ladder sum (and hence the WL effect).

The proof that the set of diagrams shown in Fig. 6a,b
satisfy $\nabla \cdot <{\underline \sigma}> = 0$ for each $n$ is as follows.
The set shown consists of a crossed ladder with
$n$ scatterings and all diagrams containing $n-j$ scatterings in
the crossed ladder and $j$ crossed lines dressing the crossed ladder on one
of the two Green's functions, for $j=1,2 \ldots n$. (For $j>1$ we have
included diagrams of lower order in $(k_fl)^{-1}$.)
It is helpful to redraw these diagrams as shown in the figures; in this way
it is clear that this set is generated
simply by moving one of the vertices $\tilde{V}$ around the figure, passing
through the impurity lines one by one.
Without loss of generality, we may take the divergence at
the ``moving'' vertex.  As just shown, taking the divergence of each diagram
generates {\it two} diagrams, one in which the vertex is
moved to the impurity line immediately to the left, the other in which the
sign is reversed and the vertex
is moved to the impurity line immediately to the right (and in both cases
the ladder dressing is removed).  This alternation in sign leads to
a cancellation between the ``right'' diagram generated from the conductivity
diagram with $j=m$ and the
the ``left'' diagram generated by the one with $j=m+1$.
This cancellation procedes around the figure until the only remaining
divergence diagram is the one where the divergence vertex
has moved all the way around the
crossed ladder and lies adjacent to the remaining dressed velocity vertex
attached at ${\bf r^{\prime} }$.  This diagram may be exactly cancelled by
adding to the
conductivity the set of diagrams shown in Fig. 6b consisting of a crossed
ladder self-energy insertion to one of the Green functions dressed by an
ordinary ladder.  The divergence of this new set cancels internally except for
the diagram shown in Fig. 6b.  This diagram is exactly the negative of
the remaining divergence diagram from the first set, except that it
doesn't contain the bare vertex.  Thus the sum of the divergence of both
sets cancels exactly, except for the diagram with the bare velocity vertex
evaluated at ${\bf r}={\bf r^{\prime} }$.  However we noted above that such a
diagram
vanishes when the antisymmetric derivative with respect to ${\bf r^{\prime} }$
is taken. Thus we have shown that the sum of the two sets
does indeed yield a current-conserving approximation to $<{\underline
\sigma}>$.

We note that this operation of ``moving a vertex around the figure" is
a quite general way of constructing a current-conserving set of diagrams
from a given diagram.  In fact, the additional ladder dressings on the
crossed ladder that were needed to complete the cancellation can be
regarded as the result of moving the vertex
through the ladder dressing on the other vertex $\tilde{V}$ (the one
attached to the point ${\bf r^{\prime} }$ in Fig. 6a.

\subsection{Limitations of Diffuson and Cooperon Approximations}

The exact ladder sum would be obtained by solving the integral equation (9),
however in most applications to disordered conductors one is
interested in the solution in the limit of long wavelengths, so one assumes
that $L({\bf r},{\bf r^{\prime} })$ is slowly-varying compared to $L_0$ and
one expands $L({\bf r_1},{\bf r^{\prime} })$ for ${\bf r_1} \approx {\bf r}$
to obtain a differential equation for $L({\bf r},{\bf r^{\prime} })$ which
has the form of a diffusion equation\cite{scot}.  For the simplest case of zero
frequency and temperature this takes the form
\begin{equation}
-D\tau \nabla^2 d({\bf r},{\bf r^{\prime} }) =
\delta({\bf r}-{\bf r^{\prime} }),
\end{equation}
where $D$ is the elastic diffusion constant, $l^2/d\tau$, and we denote
the long wavelength approximation to the ladder sum by
$d({\bf r},{\bf r^{\prime} })$.  We will henceforth refer to the exact
solution of Eq. (9) as the ladder approximation or ladder sum
while the solution to Eq. (13) will be referred to as the {\it diffuson}
approximation (the terms are often used interchangeably
in the literature).  In the absence of a magnetic field the crossed ladder
sum of Fig. 2b satisfies the same integral equation as the ladder sum,
and thus the long-wavelength approximation, which we denote as
the {\it cooperon},
satisfies the same diffusion equation.  However, in contrast to the diffuson,
the cooperon is sensitive to weak magnetic fields, which enter the diffusion
equation by the minimal substitution for particles of charge $2e$,
$i\nabla \rightarrow i\nabla - 2e{\bf A}/\hbar c$ \cite{RMP,scot}.
The appearance of the
magnetic field in this way leads to the suppression of weak localization
by a magnetic field through the elimination of the diffusion pole.

Although the diffuson/cooperon approximation to the ladder/crossed-ladder
sums is adequate for evaluating the conductivity tensor for widely
separated points ${\bf r},{\bf r^{\prime} }$, it causes a breakdown of local
current
conservation.  With appropriate care this breakdown can be shown not to
affect the value of the average transmission coefficients and hence of
the conductance, essentially because one is only evaluating the diffuson
at widely separated points.  However the diffuson/cooperon approximation
is not adequate to calculate the average reflection
coefficients (a fact which has been known in the optics literature for some
time \cite{Matt}).
The reason for this is that a major contribution to reflection involves short
trajectories of only a few mean free paths in length, for which one needs
an accurate expression for $L({\bf r},{\bf r^{\prime} })$ on the scale of $l$.
But the exact short wavelength behavior of the ladder
$L({\bf r},{\bf r^{\prime} })$
is known (for an infinite medium) \cite{jphyfr} and differs substantially
from that of the diffuson approximation.  As a result, although the
diffuson approximation allows a correct evaluation of $<T>$ it fails for
$<R>$ (the average total reflection) and hence $<T> + <R> \neq N_c$,
and the local violation of current conservation shows up as a global
violation of unitarity. We now describe this problem in some detail.

First we discuss the average total transmission, $<T>$,
which is just $<g>/(e^2/h)$ by Eq. (4).  If one starts from the
volume-averaged expression for $g$ it is well-known that
the Drude-Sommerfeld average conductance
is correctly obtained from the leading order diagram in $(k_fl)^{-1}$
within the white-noise model,
i.e. one finds $<g> = \sigma_0 A/L$ where $\sigma_0 = ne^2\tau/m$ and $A,L$
are the cross-sectional area and length of the conductor.
It is easy to show that this conductance corresponds to
an average transmission $<T> = (\pi/2)N_c l/L$ (in two dimensions).
In the volume-averaged approach, however, all the ladder diagrams
for the conductance vanish from isotropy so the result is insensitive to
the failure of the diffuson approximation at short distances.

In contrast, when evaluating conductance from the transmission coefficients
the ladder diagrams do not vanish since the conductivity tensor is only
integrated over a surface.  We have not found in the literature
confirmation that calculating
$<T> = \sum_{i,j}^{N_c} <T_{ij}>$ in the diffuson approximation yields
the correct result as well. We now confirm that it does, although only
after improving the standard approximation for the diffuson boundary
conditions.  In volume-averaged calculations the diffuson
is taken to vanish on the surface of the conductor connected to the
leads (this is supposed to correspond to the boundary condition of a
very large metallic contact).  However if this approximation is made
and if the point ${\bf r}$ is chosen at the surface,
then $<{\underline \sigma}({\bf r},{\bf r^{\prime} })>$ will
not conserve current even for ${\bf r^{\prime} }$ far in the bulk.
It can be shown however that if
the diffuson is set to zero a distance of order the mean free path
outside the interface \cite{Ishi} this property is restored
for the total flux crossing the interface.  In two dimensions
the correct distance is $\pi l/4$ outside the sample.
This modified boundary condition is required to get the correct average
transmission coefficients; to get correct reflection coefficients even this
approximation is inadequate and a new approximation is needed which
is discussed below.

The average transmission and reflection coefficients to leading order
are given by the diagram consisting of a single
diffuson between two current vertices.  To examine individual $T_{ij},R_{ij}$
we must evaluate these diagrams for a given modes $i,j$ which requires
integrating over the transverse directions ($y,y'$) with the mode wavefunctions
while staying in real-space for the current direction ($x,x'$).  This
is the method we use henceforth in the paper when evaluating diagrams
explicitly; all calculations are done in two dimensions for convenience.
Upon transforming to mode space, the differential operator at the current
vertices ${\stackrel{\leftrightarrow}{\nabla}} / 2mi$ simply is replaced by
the longitudinal velocity,
\begin{equation}
 <T_{ij}> =v_i v_j <G^+(x=0, \;i;x=L_x,\; j) G^-(x=L_x,\; j;
x=0, \; i)>
\end{equation}
where $v_i, v_j$ are velocities for the relevant modes.
We need the 1PGF which becomes
\begin{mathletters}
\begin{equation}
{\overline{G}^+}(x, \;i;x^{\prime}, \;j)=\delta_{i,j}\int (\frac{dp_x}{2\pi})
\frac{e^{ip_x(x-x')}}{\varepsilon_{f}-{\bf p}^2/2m +i/2\tau}
\end{equation}
\begin{equation}
 =-i \delta_{i,j}
e^{i p_j |x-x^{\prime}| - |x-x^{\prime}|/(2v_j \tau)}/v_j
\end{equation}
\end{mathletters}
\noindent
where ${\bf p}^2$ is the sum of squares of transverse and longitudinal
momenta and $p_j= \sqrt{2m \varepsilon_{f} - q_j^2}, q_j=(j\pi/L_y)^2$.
Finally, the propagation from a given mode at $x=0$ to a point
${\bf r^{\prime} }$ is given using this Green's function and projecting from
a mode to a y-position.  We obtain
\begin{equation}
{\overline{G}^+}(x, \; j; {\bf r^{\prime} })=
-i \frac{1}{v_j}e^{i p_j |x-x^{\prime}| - |x-x^{\prime}|/(2v_j
\tau)} \sin(q_j y^{\prime}) (2/L_y)^{1/2}
\end{equation}
where the factor of $(2/L_y)^{1/2}$ is
added for normalization.  In evaluating the diagram for $<T_{ij}>$
we note that the phase factors involving $x,x'$
cancel since the advanced and retarded Green's functions have opposite phases.
The phases coming from $y,y'$ vary on
the length scale $k_f^{-1}$, while the diffuson varies on at most the
length scale $l$.  So, we can replace the y-phase factor arising from the
square of $\sin(q_jy^{\prime})$ by its average.  One finds:
\begin{equation}
 \label{DiffInteg}
<T_{ij}>={c_i u^2} \frac{1}{L_y^2}\int_0^{L_x} \int_0^{L_x} dx dx^{\prime}
\left(v_i v_j \frac{1}{v_i^2} \frac{1}{v_j^2} e^{-x/(v_i \tau)}
e^{-x^{\prime}/(v_j \tau)}
\int dy \int dy^{\prime} d(x,y;x^{\prime},y^{\prime})
\right)
\end{equation}
where $x$ is measured from the left-hand side
of the sample and $x^{\prime}$ from the right-hand side.   The integral
of the 2d diffuson over $y,y'$ satisfies a 1d diffusion equation which
has the simple solution:
$d(x,x')= (x+\frac{\pi}{4}l)( x^{\prime}+\frac{\pi}{4}l)(2 L_y/l^2 L_x)$,
where the final factor of $L_y$ arises from the transverse integration
and we have dropped terms lower order in $(l/L_x)$.
The integral above can now be integrated by
parts; we obtain after some algebra and dropping exponentially small terms
\begin{equation}
 <T_{ij}>=\frac{2}{\pi N}\frac{l}{L_x}(\frac{v_i}{v_f}+\pi/4)
(\frac{v_i}{v_f}+\pi/4).
\end{equation}
Summing over the modes $i,j$ using the relation
$mv_j = \sqrt{p^2 - (j\pi/L_y)^2}$ yields the correct result
\begin{equation}
 \sum_{i,j}^{N_c}<T_{ij}>=N_c \frac{\pi}{2}\frac{l}{L_x}.
\end{equation}

If global current conservation were satisfied in the diffuson approximation
we should find
$\sum_{i,j}^{N_c}<R_{ij}>=N_c (1- \pi l/2 L_x) \approx N_c$.
However it is shown in the Appendix that exactly the same calculation for
$<R_{ij}>$ yields the result $\sum_{i,j}^{N_c}<R_{ij}> \approx 0.73N_c$.
As noted above, this failure of global
current-conservation arises from the fact that
reflection coefficients require consideration of the conductivity tensor
on scales of order the mean free path, and the diffuson approximation
to the ladder sum is not accurate on this scale.

\subsection{Compensating for the Diffuson Approximation}

The most elegant resolution to this problem would be to work with the
full ladder sum or some improved approximation to it, instead of the
diffuson. A partial improvement to the diffuson approximation is
given in the Appendix, however it is not sufficient to restore
current conservation to this order in $(k_fl)^{-1}$.
Although an exact evaluation of the
crossed-ladder sum (which differs from the ladder sum only in the absence
of the single scattering diagram), does exist  \cite{jphyfr} in momentum space,
it is not clear how to obtain a real space solution that can be used in a
sample {\it with boundaries}.  The corresponding real space differential
equation would be of infinite degree, complicating the specification of
boundary values.  The analogous problem in classical diffusion, essentially
the Milne problem, has been solved  \cite{mf}, but the solution appears
too complicated to be of any practical use in this context.

An alternative approach to restoring current conservation is to find an
approximation for the current vertex when connected to the ladder sum which
restores current conservation when the ladder is approximated by the
diffuson.  This was essentially the approach taken in the work of Kane et al.
 \cite{Kane1,Kane2}, where they treat the bare conductivity bubble as
proportional to $\delta ({\bf r}-{\bf r^{\prime} })$.
However calculation of the total reflection (for which the points
${\bf r},{\bf r^{\prime} }$ may coincide) requires a slightly more
careful treatment to avoid evaluating this delta function at zero.  First we
note that current conservation in the ladder approximation to
${<\underline \sigma >}$ followed from the cancellation between the term
$(1/\tau)|{\overline{G}^+} ({\bf r},{\bf r^{\prime} })|^2$ arising from
$\nabla \cdot V_0({\bf r},{\bf r^{\prime} })$ and that arising from the
divergence of the
bare vertex dressed by the ladder sum.  We can find a current-conserving
approximation when the ladder sum is replaced by the diffuson if
we can modify $V_0$ in the dressed vertex so as to maintain the result
that the divergence of the dressed vertex is
$-(1/\tau)|{\overline{G}^+} ({\bf r},{\bf r^{\prime} })|^2$.
The required modification can be determined by the following argument.
Note that the bare vertex depicted in Fig. 3
with ${\bf r_{1} }={\bf r_{2} }={\bf r^{\prime} }$ is a vector
function of ${\bf r - r^{\prime} }$
with zero curl.  Hence it may be written as
the gradient of a scalar function $F({\bf r - \bf r^{\prime} })$.  Therefore
the divergence of the dressed vertex becomes
\begin{equation}
\int {\bf dr'}{\bf dr''}\nabla^2 F({\bf r - \bf r^{\prime} })
d({\bf r^{\prime} },{\bf r^{\prime \prime} }){\overline{G}^+}
({\bf r^{\prime \prime} },{\bf r_{1} })
{\overline{G}^-} ({\bf r^{\prime \prime} },{\bf r_{2} }).
\end{equation}
Because $F$ is a function of ${\bf r}-{\bf r^{\prime} }$,
integrating by parts twice shifts the Laplacian onto the diffuson and allows
use of the basic equation
$\nabla^2 d({\bf r^{\prime \prime}}-{\bf r^{\prime} })=
-(1/D\tau) \delta ({\bf r^{\prime \prime}} -{\bf r^{\prime} })$
to give the result
\begin{equation}
-\frac{1}{D\tau}\int {\bf dr''}F({\bf r}-{\bf r^{\prime \prime} })
{\overline{G}^+} ({\bf r^{\prime \prime} },{\bf r_{1} })
{\overline{G}^-} ({\bf r^{\prime \prime} },{\bf r_{2} }).
\end{equation}
It follows that current conservation requires
$F({\bf r}-{\bf r^{\prime} })=D\delta ({\bf r} -{\bf r^{\prime} })$
or equivalently that $\nabla F = V_0 ({\bf r},{\bf r^{\prime} }) =
D \nabla \delta ({\bf r} -{\bf r^{\prime} })$.
Of course $V_0({\bf r}-{\bf r^{\prime} })$ is known explicitly; it is an
antisymmetric function
of range $l$, not of zero range.  Hence as claimed above the diffuson
approximation does not exactly conserve current.  However it is possible
to let the mean-free path tend to zero in the vertex Green functions such that
current conservation is restored; i.e. such that the vertex behaves as the
gradient of a delta function.  We note that for reflection it is not
possible to let all bare vertices approach the gradient of a delta function
because of the necessity of evaluating the vertex at
${\bf r}={\bf r^{\prime} }$; only the
vertex connected to the diffuson is taken to be of zero range.

The desired limit is achieved formally by changing the mean free path in
these Green functions from $l$ to $l^{\prime} \to 0$, rescaling
$k_f$ to $(k_f)l/ l^{\prime}$, and rescaling the Green functions by the
overall factor $l/ l^{\prime}$.
The antisymmetric derivative operator makes the vertex in direction
$i$ an odd function of $(r_i-r^{\prime}_i)$ and an even function of other
coordinates, where
$r_i$ is the $i$\,th position coordinate of the derivative operator and
$r_i^{\prime}$ is the ith position coordinate of the other point to which
the Green's functions in the vertex are connected.  It is straightforward
to check that the new vertex vanishes when the function against which it is
convoluted vanishes to zeroth and first order in any coordinate, and that
the vertex in direction $i$ does not vanish
when convoluted against a linear function of $(r_i-r_i^{\prime})$.  This
means that the vertex acts on a polynomial in the coordinates to select out
only terms linear in one coordinate and constant in the other two.
Therefore the vertex does act as the gradient of a delta function, and it is
again easy to check that the weight of the delta function is exactly the
diffusion constant, $D$, as required for current conservation.
A quick way of confirming that the weight is correct is to note that
for points on opposite sides of the sample the difference between using
a vertex of range $l$ and the new zero range vertex introduces
only exponentially small corrections in $l/L_x$.
Since we confirmed above that the diffuson
approximation with the old vertex gives exactly the correct value
for $<T>$ to leading order, it follows that the
leading behavior of the new vertex must be quantitatively correct.

Using this limiting procedure on the bare vertex allows us to calculate
the correct average total reflection probability using the
diffuson approximation (and employing a real-space formulation
analogous to Eq. (3)), i.e. we find $<R>=N_c$ to leading order in $l/L_x$.
However this procedure has significant limitations.  It only works for
evaluation of real-space diagrams or diagrams with one vertex in real-space
and one in mode space (the one not taken to be zero range).  We have not
found a sensible current-conserving approximation incorporating
the diffuson approximation which will allow us
to evaluate the individual $<R_{ij}>$, so explicit
calculations of reflection coefficient diagrams typically contain errors of
order unity in prefactors, although the dependence on parameters such as
$k_f$, $l$, and $N_c$ is correct.

\section{Effects of Timereversal-Symmetry Breaking on Reflection}

\subsection{Change in $<R>$ due to TR Symmetry}

Although individual diagrams for $<R_{ij}>$ cannot be evaluated precisely
due to the inaccuracy of the diffuson approximation, we can evaluate the
sum of all the diagrams over $i,j$ from current conservation.
Since we have proven
the diagrams of Fig. 6 form a current-conserving set we may replace each
diagram by its volume average after summing over $i,j$.  In this
case one is essentially using
the relationship $<R> = N_c - <T>$
and calculating conductance diagrams.  All diagrams
except the one with the single crossed ladder vanish; this is the
familiar diagram which upon evaluation in the cooperon approximation
to the crossed ladder gives $\delta <T>_{WL} = -1/3$  \cite{Mello}
in the limit of $L_x \gg L_y$ (quasi-1d sample).  The only question
raised by our derivation is whether one expects the cooperon approximation
to be adequate here as again the points ${\bf r},{\bf r^{\prime} }$ may
coincide, and in
fact standard treatments  \cite{scot} express the WL correction in
terms of $c({\bf r},{\bf r})$.  However the volume-averaging again appears to
rescue this approximation.  Any diagram with just a few scatterings will
be of order $l/L_x$, and may be neglected.  It is only the diagrams with
a number of impurity scatterings $n \sim (L_x/l)^2$ which are long-range
enough to give a result of order unity, and for such diagrams the
vertex may indeed be treated as short range and the cooperon approximation
is adequate.  Since a weak magnetic field
kills the cooperon contribution, the change in total reflection due to
breaking TR symmetry is $\delta <R> = <R(B)> - <R(0)> = - 1/3$.
This leads to a positive magnetoconductance which saturates at a value
$<g(B)>-<g(B=0)> \equiv \delta g_B = (+1/3)(e^2/h)$.

\subsection{Evaluation of $\delta R_{ij}$}

Although we cannot evaluate all reflection diagrams precisely, we can
accurately estimate the diagonal terms which correspond to the coherent
backscattering effect discussed in the introduction.  The simplest
diagram is just the crossed ladder with no dressings and a fixed mode
$i,j$ at each vertex.  For $i \neq j$ this vanishes using Eq. (17) due to
the orthogonality of different mode wavefunctions; for $i=j$ the diagram
is equal to the ladder sum for $<R_{ii}>$ except for the absence
of the single-scattering diagram.
Hence we have $\sum_i^{N_c} <R_{ii}>_{coop} \equiv \delta R_D \approx 1$.
Thus we see that the diagonal reflection is enhanced in
the presence of TR symmetry by roughly a factor of two,
confirming our statement
that this diagram is the mode-space manifestation of the coherent
backscattering peak.  This diagram makes no contribution to off-diagonal
reflection.  Therefore if coherent backscattering and WL were equivalent
one would expect the WL effect in conductance to be $e^2/h$ and not
$(1/3) e^2/h$ as we just found.
The other dressed diagrams for $<R_{ij}>$ are not
diagonal in mode-space and must add up to the difference $(-2/3)e^2/h$.

Unfortunately, due to the inaccuracy of the cooperon approximation there
is no reason that explicit evaluation of these diagrams will yield
precisely $(-2/3)e^2/h$.
Thus we simply show that there exist diagrams with the appropriate order
of magnitude and sign to give the expected effect.
We note that all the other diagrams we have introduced which include dressings
on the cooperon are of order $1/N_c^2$.
Thus after summing over all modes they will indeed give a
contribution of order unity.
We evaluate the simplest of the diagrams (see Fig. 8)
which yields a negative
contribution to reflection coefficients. This will illustrate the general
properties of these diagrams and also shows
that time-reversal symmetry can indeed reduce the off-diagonal reflection
(as well as enhancing the diagonal reflection).

The diagram has an overall negative sign arising from the presence of two
extra advance Green functions with no compensating retarded functions
(each 1PGF gives a factor i).  There are two such diagrams (the second is
obtained by dressing the top of the diagram) with the same value.
The cooperon is non-zero only if the transverse momentum on one side
is the same magnitude as on the other side. This does not hold exactly for the
crossed ladder, but it is still sharply peaked about reflection into the same
mode.  To lowest order in $(k_fl)^{-1}$ we have that the x-position of the
impurity dressing must be less than that of both ends of the cooperon.  The
integration over the x-position of this point then will not introduce any
additional phase.  After integrating by parts and some algebra we find
that this diagram and its conjugate contribute
to reflection from mode $i$ to mode $j$
\begin{equation}
 -4\frac{({c_i u^2})^2}{l^2 L_y^2 v_i^2 v_j^2} \left(
\frac{\pi}{4}l\frac{\tau^3v_i^2v_j^2}{v_i+v_j}+2\frac{v_i^3v_j^3\tau^4}
{(v_i+v_j)^2}
\right).
\end{equation}
After some simplifications this becomes
\begin{equation}
-\frac{1}{N_c^2}\frac{4}{\pi^2}
\left(\frac{\pi}{4}\frac{v_f}{v_i+v_j}+2\frac{v_iv_j}{(v_i+v_j)^2}
\right) .
\end{equation}
As noted above, the contribution is negative and of order $1/N_c^2$.
This diagram is typical
of the {\it off-diagonal} contributions to reflection coefficients in the
presence of time-reversal symmetry, i.e. it vanishes on volume averaging
and is of order $1/N_c^2$, but gives a total contribution to reflection
{\it of the same order as the bare crossed ladder diagram} because there is
no restriction $i=j$.

Such contributions are hard to understand from a (naive) semiclassical
point of view.  Since the diffuson and cooperon parts of such diagrams
are typically represented in real-space, it has become common in the
literature \cite{cs,Kane1,scot} to interpret each diagram as representing a
type of interference between paths.  With this interpretation the
cooperon represents a time-reversed pair of counter-propagating paths
which form a closed loop, consistent with the well-known interpretation
of the CB peak mentioned in the introduction.
However it must be kept in mind that the ``paths'' represented by such
diagrams are not classical paths as they would be in a true semiclassical
theory.  Impurity-averaged perturbation theory in real-space allows any
piece-wise linear path to contribute, and this complicates their semiclassical
interpretation.  For example, diagrams of the type shown in Fig. 6 in which
the vertices are dressed with ladder insertions
would correspond to two paths which diffuse together to the same point in
the sample, but then diverge and traverse a loop in opposite directions,
before possibly rejoining.  Such a pair of paths is impossible classically,
so the flux-sensitivity they impart to the reflection coefficients has
no straightforward semiclassical interpretation.
Although the white-noise model we have used has no simple classical limit,
and one might speculate that it is in some respects untypical,
it is noteworthy that such off-diagonal correlations are found numerically
in simple chaotic systems \cite{wl1,wl2}.  Our work makes it
clear that such contributions are essentially required by current conservation,
and to our knowledge current conservation has not been proven for
the semiclassical approximation to the scattering coefficients.  Thus we
may hope that a deeper understanding of these results will arise from
progress in the study of current conservation within the semiclassical
approximation.

\section{Symmetry Effects}

Our above analysis of the off-diagonal contribution to the WL effect in
disordered conductors was
motivated by numerical results obtained for the WL effect in ballistic
conductors with chaotic classical dynamics \cite{wl1,wl2}.
These numerical studies
also revealed another aspect of the WL effect which apparently has not
been considered previously, the sensitivity of the effect to spatial symmetry.
A particularly striking result of this work was a {\it reduction} of the
WL positive magnetoconductance
for ballistic junctions with double reflection symmetry; in
this case the magnitude of the WL obtained numerically was slightly
positive, but consistent with zero to the statistical accuracy
of the calculations.
Unlike the off-diagonal contribution to WL, which exists in both ordered and
disordered media, the symmetry-induced contributions obviously do not
exist in typical disordered media unless some very special process enforces
a symmetry in the random potential. However it is still helpful to
consider an ensemble of potentials which are random except for the presence
of discrete spatial symmetries since in this case analytic calculations
may be performed which provide insight into the origin of the sensitivity
to symmetry and also may resolve the intriguing question of whether
certain symmetries can actually eliminate or reverse the sign of the WL effect.
Thus we study the WL effect in symmetric ``random'' potentials in this section.

It is possible to consider general spatial symmetry groups using
the approach developed below. However, we find that only symmetry operations
which are their own inverse give significant contributions; thus,
we treat only single or double reflection symmetries.
For definiteness we define the x-direction to be the direction of current flow
and consider systems with reflection symmetry across the x-axis, the y-axis,
or both.  The symmetry operations on the random potential are represented
by operators $R_x,R_y$ and in the double reflection case $R_{xy}=R_xR_y$.
As before, we restrict ourselves to the two-dimensional case but
extension to higher dimensions is trivial.
The notation
${\underline{{\bf r}}}_x,{\underline{{\bf r}}}_y,{\underline{{\bf r}}}_{xy}$
will be used to represent the point related to ${\bf r}$
by the relevant symmetry operator (the subscripts will be omitted when
it is not necessary to distinguish the different symmetry operations).
We modify the standard white-noise model for
the impurity potential to include symmetry effects as follows:
$<V({\bf r})> = 0$,
$<V({\bf r})V({\bf r^{\prime} })>= c_iu^2[\delta ({\bf r} - {\bf r^{\prime} })
+ \sum_{\mu} \delta ({\underline{{\bf r}}}_{\mu} - {\bf r^{\prime} })]$,
where the index $\mu$ runs over the complete set of symmetry
operations.  As before averages of higher even powers of $V$ are given by
the sum of all pairwise averages.

With this modification the effective interaction mediated by the impurity
potential no longer conserves momentum.  If momentum ${\bf p}$ is lost by
scattering off an impurity, $\underline{{\bf p}}$
may be gained at the reflected impurity.  As a result, the 1PGF is no longer
diagonal in momentum space.  However the modifications of the 1PGF
will be negligible for our purposes.
If we compare the contribution to the self-energy from scattering off the
same impurity twice to that due to scattering from the impurity and its
reflected impurity (see Fig. 9 ), the latter is only important if the
impurity is within a few mean free paths of the reflection plane.  This
means that changes in the 1PGF produced by the new diagram
will produce changes in the conductance that, after volume averaging,
are lower order in $l/L$, where
$L$ is a typical sample dimension.  Furthermore, even for two impurities
lying near the symmetry plane one can verify that the phase of the Green
function will not vary to first order in the transverse displacement
of the impurity if the scattering is off the same impurity but
will vary to first order when the scattering is from the reflected
impurity (due to the length of the segment propagating from one to the other).
As a result, the latter diagram will be reduced by the factor
$(k_f l)^{-1/2}$.
Thus we may use the usual impurity-averaged 1PGF in evaluating
the conductance even in the presence of symmetry.

Even though the 1PGF is unchanged the conductance is modified
because new generalized ladder and crossed-ladder vertex corrections occur
due to the presence of symmetry.  An effect of this general type
is fairly obvious from the semiclassical point of view in which the WL
arises for constructive interference of paths with the same phase; in
the presence of symmetry more such paths arise.
In particular, if we interpret the real-space diagrams as representing paths,
a crossed-ladder insertion in which the lower line is the precise reflection
of the upper line (see Fig. 10)  represents paths related by TR and
reflection symmetry (or symmetries).  We refer to this as a reflected
crossed-ladder diagram and evaluate its contribution to the conductance
in the cooperon approximation.  These new diagrams arising due to symmetry
in general have different numerical values than the standard WL diagram
due both to changes in the value of the short-ranged current vertex and
to changes in the value of the relevant cooperon insertion.
We first evaluate the changes in the current vertex.
We recall the standard procedure \cite{scot} for evaluating the usual
crossed-ladder diagram (in the cooperon approximation) in real-space.
The vertices of the diagram are treated as short-ranged and replaced with
$V_0\delta ({\bf r}-{\bf r^{\prime} })$, where $V_0$ is the volume-average of
the vertex.  The diagram then becomes proportional to the integral of
$c({\bf r},{\bf r})$ over the whole sample. To evaluate the coefficient
$V_0$ for the standard diagram, note that the vertices give a factor of
\begin{equation}
{\overline{G}^+}({\bf r},{\bf r_{1} }) v_x({\bf r_{1} })
{\overline{G}^-}({\bf r_{1} },{\bf r^{\prime} })
{\overline{G}^-}({\bf r},{\bf r_{2} }) v_x({\bf r_{2} })
{\overline{G}^+}({\bf r_{2} },{\bf r^{\prime} })
\end{equation}
After reversing the order of arguments to
${\overline{G}^-}({\bf r},{\bf r_{2} })$ and
${\overline{G}^+}({\bf r_{2} },{\bf r^{\prime} })$
(which is permitted in the presence of time-reversal symmetry) and
integrating over all the coordinates
${\bf r}, {\bf r^{\prime} }, {\bf r_{1} }, {\bf r_{2} }$ this becomes
\begin{equation}
-Tr ({\overline{G}^+} v_x {\overline{G}^-} {\overline{G}^+} v_x
{\overline{G}^-}).
\end{equation}

In the reflected crossed-ladder diagram, one or more of the Green's functions
in the vertex connect to a point ${\underline{{\bf r}}}$ instead of ${\bf r}$
and similarly for ${\bf r^{\prime} }$. The new trace may then be written
\begin{equation}
-Tr ({\overline{G}^+} v_x {\overline{G}^-} R_{\mu}
{\overline{G}^+} v_x {\overline{G}^-} R_{\mu}),
\end{equation}
where $\mu = x,y,xy$.
All the operators $R_{\mu}$ commute with ${\overline{G}}$ since
${\overline{G}}({\underline{{\bf r}}},{\bf r}')=
{\overline{G}}({\bf r},{\underline{{\bf r}}}')$; however while $R_x$ will
also commute with $v_x$, $R_y$ will {\it anti-commute}.
In addition all the $R_{\mu}$ satisfy $R_{\mu}^2=1$.
Hence we may move one of the symmetry operators through the trace by
commutation or anticommutation until it is adjacent to the other at which
point the product becomes unity.  Therefore we find that
if $\mu=y,xy$ the coefficient of the new vertex is $-V_0$ whereas
if $\mu=x$ it is just $V_0$.

The new conductance diagrams involve the reflected cooperon ${\underline{c}}
({\bf r},{\underline{{\bf r}}})$.  However it is easy to see
that ${\underline{c}}=c$ because the
kernel of the integral equation defining the reflected crossed-ladder
${\overline G}^{(+)}({\bf r},{\bf r^{\prime}})
{\overline G}^{(-)}({\underline{\bf r}},{\underline{{\bf r^{\prime}}}})$
is identical to that defining the usual crossed-ladder
given the symmetry of the Green function under reflection.
Hence the only difference from the
standard diagram (other than the possible sign change of the vertex)
is that the new diagrams are proportional to the volume average of
$c({\bf r},{\underline{{\bf r}}})$ rather
than $c({\bf r},{\bf r})$.  For a rectangular sample of length $L_x$ and width
$L_y$ with the standard boundary conditions and no magnetic field we may
write the cooperon in a spectral representation where the eigenfunctions
of the diffusion equations are the normalized product of sine and cosine
functions chosen to vanish at $x=\pm L_x/2$ and to have zero derivative
at $y=\pm L_y/2$.  Thus each term in the spectral sum will have the same
value as in sum for $c({\bf r},{\bf r})$ or same magnitude but opposite sign,
depending on whether the mode or modes reflected are even or odd under
that reflection operation.  For the most relevant case, that
of a quasi-one-dimensional sample ($L_x \gg L_y$), the cooperon is
dominated by the terms with zero wavevector in the y-direction.
In the standard cooperon sum these terms are proportional to
$\sum_{n=1}^{\infty} 1/n^2$.  If the symmetry present is either
$R_y$ or $R_{xy}$ then this sum must be replaced
by $\sum_{n=1}^{\infty} (-1)^{n+1}/n^2$ due to the alternating
symmetry of the eigenfunctions in the x-direction.
The first sum is equal to $\zeta (2)= \pi^2/6$,
while the second is reduced by a factor of two.

In addition to the fully-reflected crossed-ladder diagrams there are in
principle other vertex corrections introduced by the imposition of symmetry.
For example there are partially-reflected crossed-ladder diagrams in
which some of the impurities on the lower Green function line are
reflected and some are not.  It may be shown that these
diagrams are lower order in $(k_f l)$ and $l/L$.  Reflected ladder diagrams
will give no contribution to the conductivity as can be seen by realizing that
each vertex gives a contribution proportional to $Tr(G^+ v_x G^- R)$
which clearly vanishes: if $R$ anti-commutes with x, it trivially vanishes,
while if $R$ commutes with x, it vanishes after integrating over momenta.

In summary, if $R$ reflects across the y-axis or both the
x-axis and the y-axis, for a quasi-1D sample, the diagram gives a contribution
of opposite sign and one half the magnitude as the normal WL diagram.
If $R$ reflects across the x-axis only, the diagram contributes with the same
magnitude and sign as the normal WL diagram.  If there is 4-fold symmetry
in the sample, the total WL effect is the sum of that for the non-symmetric
case, plus diagrams in which $R$ reflects across the x-axis alone, the y-
axis alone, and both axes.  As we reviewed above, the normal WL diagram
is equal to $-1/3$ (in units of $e^2/h$, neglecting spin); this gives
the difference between the
average conductance and the Drude average conductance (obtained from
the diffuson diagrams).  If we define $\delta g_{\mu}, \; \mu=x,y,xy$
as the difference between the Drude conductance and the average
conductance in the presence of {\it both} TR and the relevant spatial symmetry
our results imply: \begin{equation}
\delta g_{x}=-2/3,\;\;\;\delta g_{y}=-1/6,\;\;\;
\delta g_{xy}=-1/3.
\end{equation}
Thus we arrive at several non-trivial conclusions.  First, spatial symmetries
do not reverse the sign of the WL effect; in all cases the conductance is
reduced from its Drude value.  Second, relative to the case of no symmetry,
symmetry across the direction of current flow enhances WL, whereas symmetry
perpendicular to the direction of current flow decreases WL.  Finally,
the combination of both symmetries leads to no change in the WL.
All of these predictions can (and will) be tested numerically by
averaging the conductance over realizations of the random potential
which have the requisite symmetries and comparing them to the average without
any symmetry.  Thus we have in principle a means to ``measure'' the magnitude
of
the WL effect without imposition of a magnetic field.

On the other hand in experiments on disordered conductors
the WL effect is measured
by the difference between the conductance at B=0 and at a magnetic field
sufficient to destroy TR symmetry.  If additional spatial symmetries are
present the situation becomes more complex since in a uniform magnetic field
a combination of spatial reflection plus TR may still yield an anti-unitary
symmetry.  This fact is familiar in the context of quantum chaos where it
has been given the name ``false TR symmetry breaking'' \cite{RobnikBerry}.
In particular, a single spatial reflection (either $R_x$ or $R_y$) will
reverse the sense of circulation(see Fig. 11)
of any closed loop; since the TR operator
will again reverse the sense of circulation (in a uniform field) a closed
loop and its time-reversed, reflected partner will enclose the same flux
and not be ``dephased'' by the field.  Hence the single reflection diagrams
will not be eliminated by a uniform magnetic field and will not contribute
to the quantity $\delta g_B = <g(B)> - <g(B=0)>$, which is the experimentally
measured WL effect under normal conditions.  The
standard unreflected diagram is of course always eliminated by the field,
so we conclude that the WL effect {\it as usually measured} will be
unchanged by the imposition of single reflection symmetry and
in quasi-1D will be $\delta g_B= 1/3$.  However, double reflection symmetry
will cause a closed loop to return to its original sense of circulation,
hence TR plus double reflection will reverse the sense and such a diagram
{\it will} be eliminated by a magnetic field.
As shown above, the double reflection
diagram has one half the magnitude and opposite sign to the standard
diagram, thus the measured WL effect $\delta g_B = 1/6$ and is reduced from
its value in the absence of symmetry.  Therefore the effect of the imposition
of symmetry on the average conductance at zero field and
on the field-dependent part of the conductance $\delta g_B$
differ significantly.

These results may be tested numerically in two contexts.  First for
disordered quasi-1D conductors with random potentials chosen to have
no symmetry and then the symmetries $R_x,R_y,R_{xy}$.
Since the average conductance with no symmetry already
includes the standard WL effect
it is more convenient to compare to the predictions for a quantity
$\delta g_s \equiv <g>_{symm} - <g>_{non-symm}$.  In addition we can
compare the predictions of the theory to the numerical results obtained for
ordered cavities with chaotic classical dynamics (the results which
motivated this study).  In this case we have no microscopic analytic theory
at present. However, several studies \cite{Iida,BM,JP}
have used random matrix theory in either its Hamiltonian \cite{Iida} or
S-matrix \cite{BM,JP} form to show that the standard
(non-symmetric) WL effect is $0.25$, and this is
consistent with recent numerical results \cite{wl1,wl2,BM}.
Hence a plausible extension of our results
for quasi-1D disordered systems would rescale $\delta g_s, \delta g_B$ by
the factor $3/4$; so for example we would expect
a chaotic structure with double reflection symmetry to have
$\delta g_B = 1/8$ instead of $1/6$.  In the chaotic case the averages
are obtained by averaging over many values of the fermi energy as
described in Ref. \cite{wl1}.  Because of non-universal effects associated
with short trajectories in the chaotic case, it is more difficult to compare
two structures which are the same except for the imposition of spatial
symmetry; however we have devised a method of analyzing the numerical
results which we believe substantially overcomes this problem \cite{numan}.
The numerical results were all obtained by the recursive Green function
method \cite{recGreen,ballistic} which has been widely used
in previous studies of disordered conductors and more recently in the
study of ordered but chaotic microstructures \cite{wl2}.

Our numerical results are presented in Tables I and II and compared to the
analytic theory.  It is worth noting several features of these results.
First, the sign of the
effect is always as expected from the theory; in particular the positive
value of $\delta g_s$ for symmetry perpendicular to the current as
contrasted with the negative value for symmetry parallel is found both in
the disordered and chaotic cases.  Second, the magnitude of the effects
found numerically tend to be smaller than that predicted by the theory,
but are internally consistent.  For example the theory predicts that
$\delta g_B$ should be the same for the three cases of no symmetry,
symmetry parallel to current and symmetry perpendicular to the current.
In both the disordered and chaotic systems these three cases closely agree
even though the magnitudes found are $30-40\%$ low \cite{puzzle}.
Finally, the case
of four-fold symmetry tends to have the greatest discrepancy from the
theory, particularly in the chaotic structures.
Overall, in our view the numerical results are in good
qualitative agreement with the analytic theory and in reasonable quantitative
agreement.

\section{Summary and Conclusions}

We have presented a current-conserving approximation for the
the local conductivity tensor which includes weak localization corrections and
used it to show that the weak localization correction to the conductance
has important contributions not attributable to coherent backscattering.
The existence of such contributions, which do not have simple semiclassical
analogues, complicates the interpretation of weak localization as compared
to coherent backscattering.  Numerical results indicate that such
contributions occur both for disordered and chaotic conductors, raising
questions about the possibility of a complete semiclassical theory of
weak localization which are at present unanswered.
A related effect, relevant to weak localization in chaotic conductors,
is the sensitivity of the effect to discrete symmetries.  We showed
that for disordered conductors reflection symmetries can either
increase or decrease the weak localization effect, but cannot reverse its
sign.  Again these analytic results for disordered conductors are
consistent with numerical results from chaotic conductors.
The results relating to symmetry suggest the possibility of using the
weak localization effect as a diagnostic of the symmetry of a ballistic
microstructure.

\acknowledgements
We thank H. Bruus, S. Hershfield, and R. A. Jalabert for valuable
discusssions.  The work at Yale was supported by NSF grant DMR-9215065.

\appendix
\section{Evaluation of Reflection Coefficients Within the Diffuson
Approximation}
We evaluate the reflection coefficients within the diffuson approximation,
using the boundary conditions of no net flux entering the sample, as
discussed above.  Since the purpose of this exercise is to demonstrate that
the diffusion approximation is insufficient to handle reflection, we
calculate only the total reflection coefficients to lowest order in
$(k_fl)^{-1}$,
or, equivalently, to lowest order in $1/N_c$
where $N_c$ is the number of
transverse modes, equal to $L_yk_f/\pi$.  The diagram which we must
evaluate is shown in Fig. 7.  The simplest method of evaluating this
diagram is
to use real-space in the x-direction and momentum-space in the y-direction
as described in the text in section (3.4).  We obtain in
this case the same integral as in equation (\ref{DiffInteg}) except that now
we measure point $x^{\prime}$ from the same side of the sample as point $x$.
In this case the diffuson, integrated over transverse positions, becomes:
\begin{equation}
 d(x,x^{\prime})=
\left\{
\begin{array}{ll}
\parbox{2.5in}{ \vspace{-\abovedisplayskip} \[
\frac{2}{l^2}L_y\frac{(x+\pi l/4)(L_x+\pi l/4-x^{\prime})}{L_x+\pi l/2} ,
\] \vspace{-\belowdisplayskip} }
& \mbox{for}\; x<x^{\prime} \\
\parbox{2.5in}{ \vspace{-\abovedisplayskip} \[
\frac{2}{l^2}L_y\frac{(x^{\prime}+\pi l/4)(L_x+\pi l/4-x)}{L_x+\pi l/2} ,
\] \vspace{-\belowdisplayskip} }
& \mbox{for}\; x^{\prime}<x.
\end{array}
\right.
\end{equation}
To lowest order in $(l/L_x)$ we may replace this by
$(2/l^2)(\pi l/4+ \min(x,x^{\prime}))$.
For transmission coefficients we were guaranteed that one end of the
diffuson would be to the right of the other due to the exponential decay
of the Green's functions on the vertices; here we are not.  The integration
domain may be broken into $x<x^{\prime}$ and $x>x^{\prime}$ and
the integral may be performed by parts.  We obtain:
\begin{equation}
 <R_{ij}>=\frac{2}{\pi}\frac{1}{N}\left(\frac{\pi}{4}+\frac{v_iv_j}
{(v_i+v_j)v_f}\right).
\end{equation}
Performing the sum over $i,j$ numerically we obtain $\approx (0.73)N_c$,
instead of the value $N_c$  (up to corrections of order $l/L$) as required
by current conservation and the known value of the average conductance.

We know that calculating these diagrams using the correct ladder sum
(instead of the diffuson approximation to it) will give the correct answer
as we have proved that the ladder approximation conserves current.
Since the exact ladder sum is not analytically tractable, we instead
improve on the diffuson approximation
by explicitly evaluating the single
scattering term in the ladder sum,
and then evaluating a diagram with a single scattering dressing
a diffuson.
Thus the single-scattering contribution is
evaluated exactly, and all double and higher scattering contributions are
evaluated in the diffuson approximation.  The single
scattering can be done simply, giving a total contribution of
$\approx 0.21N_c$.
The diffuson approximation for the second and higher scatterings is more
tedious to evaluate but is found to sum to $\approx. 63N_c$.
Hence the total average reflection in this improved approximation
has increased to $\approx 0.84N_c$.  In principle this method could be
extended to even higher scatterings and an even better approximation
for the average reflection will be obtained.

\begin{figure}
\caption{
Self-consistent Born approximation (SCBA)
to the average one particle Green's
function.  Dashed lines represent impurity scatterings, thin lines represent
bare Green's functions, thick lines represent Green's functions after
averaging.
}
\end{figure}

\begin{figure}
\caption{
Expansion for ladder sum. b) Expansion for crossed ladder sum.  The +
and - signs denote advanced and retarded Green's
functions, L and X the ladder and crossed ladder sums.
}
\end{figure}

\begin{figure}
\caption{
Divergence of a general vertex in a conductivity diagram.  The arrow
represents the operation of taking the divergence, the wavy line at
the vertex represents a current operator and the solid dot
indicates a current vertex deleted by the divergence operation as
described in the text.
}
\end{figure}

\begin{figure}
\caption{
a) Divergence of a vertex dressed with impurity scatterings. b)
Divergence of bare vertex plus vertex dressed with ladder sum, represented
by $\tilde{V}$.
}
\end{figure}

\begin{figure}
\caption{
The ladder approximation to conductivity.
}
\end{figure}

\begin{figure}
\caption{
Current conserving set of diagrams containing a crossed ladder with n
scatterings.  The triangular vertex denotes the dressed vertex $\tilde{V}$.
a)The first set of diagrams as described in the text. b)
The second set as described in the text.
}
\end{figure}

\begin{figure}
\caption{
Diffuson diagram used to calculate contributions to the
transmission coefficients $T_{ij}$ to lowest order in $1/N_c$.
}
\end{figure}

\begin{figure}
\caption{ The simplest negative contribution to $R_{ij}$ due to time
reversal symmetry, evaluated in the cooperon approximation.
}
\end{figure}

\begin{figure}
\caption{
The SCBA in the presence of symmetry.  The corrections to the SCBA
self-energy due to symmetry (last diagram) are found to be negligible.
}
\end{figure}

\begin{figure}
\caption{
Crossed ladder diagram with one side spatially reflected, denoted by the
symbol ${\underline X}$.
}
\end{figure}

\begin{figure}
\caption{
Effect of reflection on sense of a path.  Single reflection reverses sense,
double reflection preserves it.
}
\end{figure}

\begin{figure}
\caption{
Improved form of the diffuson approximation as discussed in the Appendix.
}
\end{figure}

\begin{table}
\caption{
The effect of reflection symmetry on weak localization in quasi-1D
disordered conductors. The results of numerical calculation are in good
qualitative agreement with the analytic theory. The numerical results
were obtained for the standard model \protect\cite{recGreen,ballistic} of a
tight-binding Hamiltonian with on-site disorder.
A rectangular strip with the following parameters was used:
length/width $=8$, width=120, disorder strength =0.9, $k_f a =1.61$,
mean-free-path=25, $BA/\phi_0 =$ 0 or 10, and number of configurations=300.
}
\vspace{2ex}
\begin {tabular} {ldddd}
\multicolumn{5}{c}
{Symmetry Effects on Weak Localization: Quasi-1D Disordered Conductors} \\
\hline
\multicolumn{1}{c}{Symmetry Type} &
\multicolumn{1}{c}{$\delta g_s$ (theory)} &
\multicolumn{1}{c}{$\delta g_s$ (num.)} &
\multicolumn{1}{c}{$\delta g_B$ (theory)} &
\multicolumn{1}{c}{$\delta g_B$ (num.)} \\
\hline
No Symmetry &
0 & \multicolumn{1}{c}{---} & 0.333 & 0.237 $\pm$ 0.026 \\

Symmetry $\parallel$ to Current ($R_x$) &
$-$0.333 & $-$0.272 $\pm$ 0.037 & 0.333 & 0.211 $\pm$ 0.036 \\

Symmetry $\perp$ to Current ($R_y$) &
$+$0.167 & 0.129 $\pm$ 0.037 & 0.333 & 0.273 $\pm$ 0.036 \\

4-fold Symmetry ($R_xR_y$) &
0 & 0.031 $\pm$ 0.051 & 0.167 & 0.059 $\pm$ 0.059 \\
\end {tabular}

\end{table}

\begin{table}
\caption{
The effect of reflection symmetry on weak localization in chaotic
ballistic conductors. The theoretical results are obtained by scaling
the results for the disordered conductors by a factor of $3/4$ while the
numerical results are obtained for stadium-like billiards, the billiard in
Fig. 1 of Ref. \protect\cite{wl1} for the asymmetric case and its symmetry
generated analogs. For the numerical results, the average
\protect\cite{numan} was taken
over the energy range $k_f W/\pi \in$ $[4,25]$ ($[4,15]$) for the structures
with (without) $R_x$ symmetry, the magnetic field used was
$BA/\phi_0 =$ 0 or 2, and $k_f a \approx 1.0$.
Note the good qualitative agreement except in the four-fold symmetric case.
}
\vspace{2ex}
\begin {tabular} {ldddd}
\multicolumn{5}{c}
{Symmetry Effects on Weak Localization: Chaotic Ballistic Conductors} \\
\hline
\multicolumn{1}{c}{Symmetry Type} &
\multicolumn{1}{c}{$\delta g_s$ (theory)} &
\multicolumn{1}{c}{$\delta g_s$ (num.)} &
\multicolumn{1}{c}{$\delta g_B$ (theory)} &
\multicolumn{1}{c}{$\delta g_B$ (num.)} \\
\hline
No Symmetry &
0 & \multicolumn{1}{c}{---} & 0.25 & 0.147 $\pm$ 0.018 \\

Symmetry $\parallel$ to Current ($R_x$) &
$-$0.25 & $-$0.245 $\pm$ 0.038 & 0.25 & 0.148 $\pm$ 0.020 \\

Symmetry $\perp$ to Current ($R_y$) &
$+$0.125 & 0.170 $\pm$ 0.030 & 0.25 & 0.161 $\pm$ 0.019 \\

4-fold Symmetry ($R_xR_y$) &
0 & 0.229 $\pm$ 0.043 & 0.125 & 0.043 $\pm$ 0.031 \\
\end {tabular}

\end{table}


\begin{references}
\bibitem{numan}
In the chaotic ballistic case, the numerical analysis was complicated by
the fact that changing the symmetry of the structure introduced a small
but significant change in the classical transmission. In order to obtain
results not influenced by the change in the classical transmission, we
adopted the following procedure. (1) Fit a straight line to $T(k)$. The
slope, denoted ${\cal T}$, has been previously shown to be the classical
transmission probability \cite{wl1}. (2) Subtract the classical transmission,
$\delta T(k) \equiv T(k) - {\cal T}kW/\pi$. (3) Average $\delta T(k)$
over $k$ for the different symmetry classes and compare them.
\bibitem{puzzle}
One can think of many reasons that the agreement between numerics
and analytic theory is imperfect by this amount for the chaotic conductors.
For example, in this case our argument for neglecting the change in the
1PGF due to symmetry could fail as the size of the structure effectively
sets the mean-free path.  However it is puzzling that the results for the
disordered case are consistently low by such an amount.  There
are certain parametric inequalities such as $l \ll L_x,L_y$ which are
assumed in the analytic theory but are hard to satisfy strongly in
the numerical simulations.  Perhaps these account for the discrepancy;
however in the case of the conductance fluctuations, which is similar
in many respects, one finds agreement between analytic theory and
numerics to $5\%$ accuracy with little effort.
\end{references}
\end{document}